\documentstyle[12pt]{article}
\begin{document}

\titlepage
\def\thefootnote{\fnsymbol{footnote}}

{\it University of Shizuoka}

\hspace*{11cm} {\bf US-95-01}\\[-.15in]

\hspace*{11cm} {\bf January 1995}\\[.4in]

\begin{center}

{\large\bf New Physics from }\\[.1in]
{\large\bf U(3)-Family Nonet Higgs Boson Scenario\footnote{
Talk presented at the INS Workshop ``Physics of $e^+e^-$, $e^-\gamma$ and
$\gamma\gamma$ collisions at linear accelerators", INS,
University of Tokyo, December 20 -- 22, 1995.}
}\\[.2in]

{\bf Yoshio Koide\footnote{
E-mail address: koide@u-shizuoka-ken.ac.jp}} \\[.1in]

Department of Physics, University of Shizuoka \\[.1in]
52-1 Yada, Shizuoka 422, Japan \\[.5in]

{\large\bf Abstract}\\[.1in]
\end{center}
\begin{quotation}
Being inspired by a phenomenological success of a charged lepton mass formula,
a model with U(3)-family nonet Higgs bosons is proposed.
Here, the Higgs bosons $\phi_L$ ($\phi_R$) couple only between light
fermions (quarks and leptons) $f_L$ ($f_R$) and super-heavy vector-like
fermions $F_R$ ($F_L$), so that the model leads to
a seesaw-type mass matrix $M_f\simeq m_L M_F^{-1} m_R$ for quarks and
leptons $f=u,d,\nu$ and $e$.
Lower bounds of the physical Higgs boson masses  are deduced from the present
experimental data and possible new physics from the present scenario
is speculated.
\end{quotation}

%
%--------------------------------------------------------------------
\title{New Physics from U(3)-Family Nonet Higgs Boson Scenario}
\author{Yoshio KOIDE\thanks{E-mail: koide@u-shizuoka-ken.ac.jp}\\
Department of Physics,\ University of Shizuoka\\
52-1 Yada,\ Shizuoka 422, JAPAN} \date{ }
\maketitle
\begin{abstract}
Being inspired by a phenomenological success of a charged lepton mass formula,
a model with U(3)-family nonet Higgs bosons is proposed.
Here, the Higgs bosons $\phi_L$ ($\phi_R$) couple only between light
fermions (quarks and leptons) $f_L$ ($f_R$) and super-heavy vector-like
fermions $F_R$ ($F_L$), so that the model leads to
a seesaw-type mass matrix $M_f\simeq m_L M_F^{-1} m_R$ for quarks and
leptons $f=u,d,\nu$ and $e$.
Lower bounds of the physical Higgs boson masses  are deduced from the present
experimental data and possible new physics from the present scenario
is speculated.

\end{abstract}
%--------------------------------------------------------------------
\renewcommand{\thefootnote}{\sharp\arabic{footnote}}
\renewcommand{\theequation}{\arabic{section}.\arabic{equation}}
%--------------------------------------------------------------------
%\baselineskip=21.0pt plus 0.2pt minus 0.1pt
%--------------------------------------------------------------------
\setcounter{footnote}{0}

\section{Motives}

One of my dissatisfactions with the standard model is that for the
explanation of the mass spectra of quarks and leptons,
we must choose the coefficients $y_{ij}^f$ in the Yukawa coupling
$\sum_f \sum_{i,j} \overline{f}_L^i f_{jR} \langle\phi^0\rangle$
($f=\nu,e,u,d$, and $i,j$ are family indices) ``by hand".
In order to reduce this dissatisfaction, for example, let us suppose
U(3)$_{family}$ nonet Higgs fields
which couple with fermions as
$\sum_f\sum_{i,j}\overline{f}_L^i\langle\phi_i^{0j}\rangle f_{jR}$.
Unfortunately, we know that the mass spectra of up- and
down-quarks and charged leptons are not identical and the Kobayashi-Maskawa
[1] (KM) matrix is not a unit matrix.
Moreover, we know that in such multi-Higgs models, in general,
flavor changing neutral currents (FCNC) appear unfavorably.

Nevertheless, I would like to dare to challenge to a model
with U(3)$_{family}$ nonet Higgs bosons
which leads to a seesaw-type quark and lepton mass matrix
$$
M_f \simeq m_L M_F^{-1} m_R  \ .\eqno(1)
$$
My motives are as follows.

One of the motives is a phenomenological success of a charged lepton
mass relation [2]
$$
m_e+m_\mu+m_\tau=\frac{2}{3}(\sqrt{m_e}+\sqrt{m_\mu}+\sqrt{m_\tau})^2
\ , \eqno(2)
$$
which predicts $m_\tau = 1776.96927\pm 0.00052\pm 0.00005$ MeV
for the input values [3] of $m_e=0.51099906\pm 0.00000015$ MeV and
$m_\mu=105.658389\pm 0.000034$ MeV (the first and second errors in (1.2)
come from the errors of $m_\mu$ and $m_e$, respectively).
Recent measurements [4] of tau lepton mass
$m_\tau = 1776.96^{+0.18+0.20}_{-0.19-0.16}$ MeV
excellently satisfies the charged lepton mass relation (2).
An attempt to derive the mass relation (2) from a Higgs model has been
tried [5]:
We assumed  U(3)$_{family}$ nonet Higgs bosons $\phi_i^j$ $(i,j=1,2,3)$,
whose potential
is given by
$$V(\phi)=\mu^2{\rm Tr}(\phi\phi^\dagger)
+\frac{1}{2}\lambda\left[{\rm Tr}(\phi\phi^\dagger)\right]^2
+\eta\phi_s\phi_s^*{\rm Tr}(\phi_{oct}\phi_{oct}^\dagger) \ . \eqno(3)
$$
Here, for simplicity, the SU(2)$_L$ structure of $\phi$ has been neglected,
and we have expressed the nonet Higgs bosons $\phi_i^j$ by
the form of $3\times 3$ matrix,
$$
 \phi=\phi_{oct}+\frac{1}{\sqrt{3}}\phi_s\; {\bf 1}\ ,\eqno(4)
$$
where $\phi_{oct}$ is the octet part of $\phi$, i.e., Tr$(\phi_{oct})=0$,
and {\bf 1} is a $3\times 3$ unit matrix.
For $\mu^2<0$, conditions for minimizing the potential (3) lead to
the relation
$$
v_s^*v_s = {\rm Tr}\left( v_{oct}^\dagger v_{oct} \right)\ ,  \eqno(5)
$$
together with $v=v^\dagger$, where $v=\langle \phi\rangle$,
$v_{oct}=\langle \phi_{oct}\rangle$ and $v_s=\langle \phi_s\rangle$,
so that we obtain the relation
$$
{\rm Tr}\left(v^2\right) =
\frac{2}{3} \left[{\rm Tr}(v)\right]^2 \ . \eqno(6)
$$
If we assume a seesaw-like mechanism for charged lepton mass matrix
$M_e$, $ M_e \simeq m M_E^{-1} m $, with $m\propto v$ and heavy lepton
mass matrix $M_E\propto {\bf 1}$, we can obtain the mass relation (2).

Another motives is a phenomenological success [6] of
quark mass matrices with a seesaw-type form (1),
where
$$
m_L\propto m_R \propto M_e^{1/2}\equiv \left(
\begin{array}{ccc}
\sqrt{m_e} & 0 & 0 \\
0 & \sqrt{m_\mu} & 0 \\
0 & 0 & \sqrt{m_\tau}
\end{array} \right) \ ,\eqno(7)
$$
$$
M_F\propto {\bf 1}+b_F e^{i\beta_F} 3X
\equiv \left(
\begin{array}{ccc}
1 & 0 & 0 \\
0 & 1 & 0 \\
0 & 0 & 1
\end{array} \right) + b_F e^{i\beta_F} \left(
\begin{array}{ccc}
1 & 1 & 1 \\
1 & 1 & 1 \\
1 & 1 & 1
\end{array} \right) \ .  \eqno(8)
$$
The model can successfully provide predictions for quark mass ratios
(not only the ratios $m_u/m_c$, $m_c/m_t$, $m_d/m_s$ and $m_s/m_b$,
but also $m_u/m_d$, $m_c/m_s$ and $m_t/m_b$) and
KM matrix parameters.

These phenomenological successes can be  reasons why
the model with a U(3)$_{family}$ nonet Higgs bosons, which
leads to a seesaw-type mass matrix (1), should be taken seriously.

%%%%%%%%%%%%%%%%%%%%%%%%%%%%%%%%%%%%%%%%%%%%%%% 2.

\section{Outline of the model}

The model is based on
SU(2)$_L\times$ SU(2)$_R\times$U(1)$_Y\times$U(3)$_{family}$ [7] symmetries.
These symmetries except for U(3)$_{family}$ are gauged.
The prototype of this model was investigated by Fusaoka and the author [8].
However, their Higgs potential leads to massless
physical Higgs bosons, so that it brings some troubles into the theory.
In the present model, the global symmetry U(3)$_{family}$ will be
broken explicitly, and not spontaneously, so that massless physical
Higgs bosons will not appear.

The quantum numbers of our fermions and Higgs bosons are summarized
in Table I.

\begin{center}
Table I. Quantum numbers of fermions and Higgs bosons

\begin{tabular}[t]{|c|c|c|c|c|} \hline\hline
      &  $Y$ & SU(2)$_L$    & SU(2)$_R$  & U(3)$_{family}$  \\
 \hline
 $f_L$  & $(\nu, \: e)_L^{Y=-1}$, $(u, \: d)_L^{Y=1/3}$
& {\bf 2} & {\bf 1} & {\bf 3}  \\
 $f_R$ & $(\nu, \: e)_R^{Y=-1}$, $(u, \: d)_R^{Y=1/3}$
& {\bf 1} & {\bf 2} & {\bf 3}   \\
$F_L$ & $N_L^{Y=0}$, $E_L^{Y=-2}$, $U_L^{Y=4/3}$, $D_L^{Y=-2/3}$
& {\bf 1} & {\bf 1} & {\bf 3}   \\
$F_R$ & $N_R^{Y=0}$, $E_R^{Y=-2}$, $U_R^{Y=4/3}$, $D_R^{Y=-2/3}$
& {\bf 1} & {\bf 1} & {\bf 3} \\  \hline
$\phi_L$ & $(\phi^+,\phi^0)_L^{Y=1}$ & {\bf 2} & {\bf 1} & {\bf 8}+{\bf 1} \\
$\phi_R$ & $(\phi^+,\phi^0)_R^{Y=1}$ &{\bf 1} & {\bf 2} & {\bf 8}+{\bf 1} \\
$\Phi_F$ & $\Phi_0^{Y=0}$, $\Phi_X^{Y=0}$ & {\bf 1} & {\bf 1} & {\bf 1},
\ {\bf 8}  \\ \hline\hline
\end{tabular}
\end{center}

%%%%%%%%%%%%%%%%%%%%%%%%%%%%%%%%%
\vspace{0.5cm}

Note that in our model there is no Higgs boson which belongs to ({\bf 2},
{\bf 2}) of SU(2)$_L\times$SU(2)$_R$.
This guarantees that we obtain a seesaw-type mass matrix (2) by
diagonalization of a $6\times 6$ mass matrix for fermions $(f,F)$:
$$
\left( \begin{array}{cc}
0 & m_L \\
m_R & M_F
\end{array} \right) \Longrightarrow
\left( \begin{array}{cc}
M_f & 0 \\
0 & M'_F
\end{array} \right) \ , \eqno(9)
$$
where
$M_f\simeq -m_L M_F^{-1}m_R$ and
$M'_F\simeq M_F$
for $M_F\gg m_L \ , \ m_R$.
(See Fig.~1.)

%%%%%%%%%%%%%%% Fig.1
\vspace*{1cm}

\setlength{\unitlength}{0.7mm}
\begin{picture}(160,100)(0,-25)
\linethickness{0.5mm}
\thicklines\put(0,30){\line(1,0){160}}
\thicklines\multiput(20,30)(40,0){4}{\vector(1,0){0}}
\multiput(40,30)(0,2){10}{\line(0,1){1}}
\multiput(80,30)(0,2){10}{\line(0,1){1}}
\multiput(120,30)(0,2){10}{\line(0,1){1}}
\multiput(40,30)(40,0){3}{\circle*{2}}
\put(0,20){\makebox(40,10)[b]{$f_L$}}
\put(40,20){\makebox(40,10)[b]{$F_R$}}
\put(80,20){\makebox(40,10)[b]{$F_L$}}
\put(120,20){\makebox(40,10)[b]{$f_R$}}
\put(30,20){\makebox(20,10)[c]{$g_f^L$}}
\put(70,20){\makebox(20,10)[c]{$g_F$}}
\put(110,20){\makebox(20,10)[c]{$g_f^R$}}
\put(0,10){\makebox(40,10)[b]{({\bf 2, 1, 3})}}
\put(40,10){\makebox(40,10)[b]{({\bf 1, 1, 3})}}
\put(80,10){\makebox(40,10)[b]{({\bf 1, 1, 3})}}
\put(120,10){\makebox(40,10)[b]{({\bf 1, 2, 3})}}

%\put(5,55){\makebox(40,10)[t]{$\langle\phi_L^0\rangle\equiv m_L/g_f^L$}}
%\put(60,55){\makebox(40,10)[t]{$\langle\Phi_F\rangle\equiv M_F/g_F$}}
%\put(115,55){\makebox(40,10)[t]{$\langle\phi_R^0\rangle\equiv m_R/g_f^R$}}
\put(20,55){\makebox(40,10)[t]{$\langle\phi_L^0\rangle$}}
\put(60,55){\makebox(40,10)[t]{$\langle\Phi_F\rangle$}}
\put(100,55){\makebox(40,10)[t]{$\langle\phi_R^0\rangle$}}
\put(15,65){\makebox(40,10)[t]{({\bf 2, 1, 8+1})}}
\put(60,65){\makebox(40,10)[t]{({\bf 1, 1, 1})}}
\put(105,65){\makebox(40,10)[t]{({\bf 1, 2, 8+1})}}
%\multiput(40,60)(40,0){3}{\line(1,1){2.5}}
%\multiput(40,60)(40,0){3}{\line(-1,-1){2.5}}
%\multiput(40,60)(40,0){3}{\line(1,-1){2.5}}
%\multiput(40,60)(40,0){3}{\line(-1,1){2.5}}
\multiput(30,40)(40,0){3}{\makebox(20,20){{\bf $\times$}}}
\end{picture}
\vspace*{-20mm}
\begin{center}
Fig.~1. \ Mass generation mechanism of $M_f\simeq m_L M_F^{-1} m_R$.
\end{center}

%%%%%%%%%%%%%%%%%%%%%%%%%%%%%%%%%%%%%%%%%%%%%%%%%%%%%

\section{Higgs potential and ``nonet" ansatz}

We assume that $\langle\phi_R\rangle \propto \langle\phi_L\rangle$,
i.e., each term in $V(\phi_R)$ takes the coefficient
which is exactly proportional to the corresponding term in $V(\phi_L)$.
This assumption means that there is a kind of ``conspiracy" between
$V(\phi_R)$ and $V(\phi_L)$.
However, in the present stage, we will not go into this problem moreover.
Hereafter, we will drop the index $L$ in $\phi_L$.

The potential $V(\phi)$ is given by
$$
V(\phi)=V_{nonet} + V_{Oct\cdot Singl} + V_{SB} \ , \eqno(10)
$$
where
$V_{nonet}$ is a part of $V(\phi)$ which satisfies a ``nonet" ansatz
stated below, $V_{Oct\cdot Singl}$ is a part which violates the ``nonet"
ansatz, and $V_{SB}$ is a term which breaks U(3)$_{family}$ explicitly.

The ``nonet" ansatz is as follows: the octet component $\phi_{oct}$
and singlet component $\phi_s$ of the Higgs scalar fields $\phi_L$
($\phi_R$) always appear with the combination of (4) in the
Lagrangian.
Under the ``nonet" ansatz, the SU(2)$_L$ invariant (and also U(3)$_{family}$
invariant) potential $V_{nonet}$ is, in general, given by
$$
V_{nonet}= \mu^2 {\rm Tr}(\overline{\phi}\phi)
+\frac{1}{2}\lambda_1
(\overline{\phi}_i^j \phi_j^i)(\overline{\phi}_k^l \phi_l^k)
$$
$$
+\frac{1}{2}\lambda_2
(\overline{\phi}_i^j \phi_k^l)(\overline{\phi}_l^k \phi_j^i)
+\frac{1}{2}\lambda_3
(\overline{\phi}_i^j \phi_k^l)(\overline{\phi}_j^i \phi_l^k)
+\frac{1}{2}\lambda_4
(\overline{\phi}_i^j \phi_j^k)(\overline{\phi}_k^l \phi_l^i)
$$
$$
+\frac{1}{2}\lambda_5
(\overline{\phi}_i^j \phi_l^i)(\overline{\phi}_k^l \phi_j^k)
+\frac{1}{2}\lambda_6
(\overline{\phi}_i^j \phi_j^k)(\overline{\phi}_l^i \phi_k^l)
+\frac{1}{2}\lambda_7
(\overline{\phi}_i^j \phi_k^l)(\overline{\phi}_j^k \phi_l^i) \ ,\eqno(11)
$$
where $(\overline{\phi} \phi)=\phi^-\phi^+
+\overline{\phi}^0 \phi^0$.

On the other hand, the ``nonet ansatz" violation terms $V_{Oct\cdot Singl}$
are given by
$$
V_{Oct\cdot Singl}=
\eta_1(\overline{\phi}_s\phi_s){\rm Tr}(\overline{\phi}_{oct} \phi_{oct})
+ \eta_2\left( \overline{\phi}_s(\phi_{oct})_i^j\right)
\left( (\overline{\phi}_{oct})_j^i \phi_s\right)
$$
$$
+ \eta_3\left( \overline{\phi}_s(\phi_{oct})_i^j\right)
\left( \overline{\phi}_s (\phi_{oct})_j^i\right)
+ \eta_3^*\left( (\overline{\phi}_{oct})_i^j \phi_s\right)
\left( (\overline{\phi}_{oct})_j^i \phi_s\right) \ .\eqno(12)
$$

For a time, we neglect the term $V_{SB}$ in (10).
For $\mu^2<0$, conditions for minimizing the potential (10) lead to
the relation
$$
v_s^2={\rm Tr}(v_{oct}^2)=\frac{-\mu^2}
{2(\lambda_1+\lambda_2+\lambda_3)+(\eta_1+\eta_2+2\eta_3)} \ , \eqno(13)
$$
under the conditions $\lambda_4+\lambda_5+2(\lambda_6+\lambda_7)=0$,
and $v=v^\dagger$, where we have put $\eta_3=\eta_3^*$ for simplicity.

Hereafter, we choose the family basis as
$$
v=\left(
\begin{array}{ccc}
v_1 & 0 & 0 \\
0 & v_2 & 0 \\
0 & 0 & v_3
\end{array}
\right)  \ .\eqno(14)
$$
For convenience, we define the parameters $z_i$ as
$$ z_i\equiv \frac{v_i}{v_0} =\sqrt{\frac{m_i^e}{m_e+m_{\mu}+m_{\tau}}} \ ,
\eqno(15)
 $$
where
$$
 v_0=(v_1^2+v_2^2+v_3^2)^{{1}/{2}} \ , \eqno(16)
$$
so that $(z_1, z_2,z_3)=(0.016473,0.23687,0.97140)$.

We define two independent diagonal elements of $\phi_{oct}$ as
$$
\begin{array}{l}
\phi_x =x_1\phi_1^1+x_2\phi_2^2+x_3\phi_3^3 \ , \\
\phi_y =y_1\phi_1^1+y_2\phi_2^2+y_3\phi_3^3 \ ,
\end{array} \eqno(17)
$$
where the coefficients $x_i$ and $y_i$ are given by
$$ x_i={\sqrt{2}} z_i -{1}/{\sqrt{3}} \ ,  \eqno(18) $$
$$
(y_1,y_2,y_3)=({x_2-x_3},{x_3-x_1},x_1-x_2)/{\sqrt{3}} \ . \eqno(19)
$$
Then, the replacement $\phi^0\rightarrow\phi^0 + v$ means that
$\phi^0_s  \rightarrow\phi^0_s + v_s$;
$\phi^0_x  \rightarrow\phi^0_x + v_x$;
$\phi^0_y  \rightarrow\phi^0_y$;
$(\phi^0)_i^j  \rightarrow(\phi^0)_i^j \ \ (i\neq j)$,
where $v_i=v_s/\sqrt{3}+x_i v_x$.
This means that even if we add a term
$$
V_{SB}=\xi \left(\overline{\phi}_y\phi_y
+\sum_{i\neq j}\overline{\phi}_i^j\phi_j^i\right) \ , \eqno(20)
$$
in the potential $V_{nonet}+V_{Oct\cdot Singl}$, the relation (13)
are still unchanged.

%%%%%%%%%%%%%%%%%%%%%%%%%%%%%%%%%%%%%%%%%%%%%%%%%%%%%%%%%%%% 4.

\section{Physical Higgs boson masses}

For convenience, we define:
$$
\begin{array}{ccc}
\left(
\begin{array}{c}
\phi^+ \\
\phi^0
\end{array} \right)
& = & \frac{1}{\sqrt{2}}
\left(
\begin{array}{c}
i\sqrt{2} \, \chi^+ \\
H^0-i\chi^0
\end{array} \right)
\end{array}
\ , \eqno(21)
$$
$$
\begin{array}{cccc}
\left(
\begin{array}{c}
\phi_1 \\
\phi_2 \\
\phi_3
\end{array} \right)
& = &
\left(
\begin{array}{ccc}
z_1 & z_2 & z_3 \\
z_1-\sqrt{\frac{2}{3}} & z_2-\sqrt{\frac{2}{3}} & z_3-\sqrt{\frac{2}{3}} \\
\sqrt{\frac{2}{3}}(z_2-z_3) & \sqrt{\frac{2}{3}}(z_3-z_1)
& \sqrt{\frac{2}{3}}(z_1-z_2)
\end{array} \right)
\left(
\begin{array}{c}
\phi_1^1 \\
\phi_2^2 \\
\phi_3^3
\end{array} \right)
\end{array}
\ . \eqno(22)
$$
Then, we obtain masses of these Higgs bosons which are sumalized in Table II.

\vglue.1in

\begin{quotation}
Table II. Higgs boson masses squared in unit of $v_0^2=(174\; {\rm GeV})^2$,
where $\overline{\xi}=\xi/v_0^2$.
For simplicity, the case of $\lambda_4=\lambda_5=\lambda_6=\lambda_7=0$
are tabled.
\end{quotation}

$$
\begin{array}{|c|c|c|c|}\hline\hline
\phi & H^0  & \chi^0 & \chi^\pm \\ \hline
m^2(\phi_1) &
\begin{array}{c}
2(\lambda_1+\lambda_2+\lambda_3)\\
+\eta_1+\eta_2+2\eta_3 \end{array}
& 0 & 0 \\ \hline
m^2(\phi_2) & -(\eta_1+\eta_2+2\eta_3) & -2(\lambda_3+2\eta_3) &
-(\lambda_2+\lambda_3 +\eta_2+2\eta_3)  \\[.1in] \hline
m^2(\phi_3) & \overline{\xi} & \overline{\xi}-2(\lambda_3+\eta_3) &
\overline{\xi}-[\lambda_2+\lambda_3 +\frac{1}{2}(\eta_2+2\eta_3)]
 \\[.1in] \hline
m^2(\phi_i^j) & =m^2(H_3^0) & =m^2(\chi_3^0) & = m^2(\chi_3^\pm)
\\[.1in] \hline\hline
\end{array}
$$

\vglue.1in

The massless states $\chi_1^\pm$ and $\chi_1^0$ are eaten by weak bosons
$W^\pm$ and $Z^0$, so that they are not physical bosons.
The mass of $W^\pm$ is given by $m^2_W=g^2 v_0^2/2$, so that
the value of $v_0$ defined by (16) is $v_0=174$ GeV.

%%%%%%%%%%%%%%%%%%%%%%%%%%%%%%%%%%%%%%%%%%%%%%%%%%%%%% 5.

\section{Interactions of the Higgs bosons}

{\bf (A) \ Interactions with gauge bosons}

Interactions of $\phi_L$ with gauge bosons are calculated from the kinetic
term Tr($D_{\mu}\overline{\phi}_{L}D^{\mu}\phi_L$). The results are as
follows :
$$
H_{EW}=+i\left( eA_{\mu}+\frac{1}{2}g_z\cos2\theta_W Z_{\mu}\right)
{\rm Tr}(\chi^{-}\buildrel\leftrightarrow\over\partial\,^{\mu} \chi^+)
+\frac{1}{2}g_z Z_{\mu}{\rm Tr}
(\chi^{0} \buildrel\leftrightarrow\over\partial\,^{\mu} H^0)
$$
$$
+ \frac{1}{2}g \left\{ W_{\mu}^+[{\rm Tr}
(\chi^{-} \buildrel\leftrightarrow\over\partial\,^{\mu} H^0)
-i(\chi^{-} \buildrel\leftrightarrow\over\partial\,^{\mu} \chi^+)
+{\rm h.c.}\right\}
$$
$$
+\frac{1}{2}\left( 2gm_W W_{\mu}^-W^{+\mu}
+g_z m_Z Z_{\mu}Z^{\mu}\right) H_1^0
\ , \eqno(23)
$$
where $g_z=g/\cos\theta_W$ and $\chi_1^{\pm}=\chi_1^0=0$.

Note that the interactions of $H^0_1$ are exactly same as that of $H^0$
in the standard model.

\vglue.1in

{\bf (B) \ Three-body interactions among Higgs bosons}

$$
H_{\phi\phi\phi}=\frac{1}{2\sqrt{2}}\frac{m^2(H_1^0)}{v_0} H_1^0
{\rm Tr}(H^0 H^0)
+\frac{1}{2\sqrt{2}}\frac{m^2(H_2^0)}{v_0}\left( H_1^0 H_2^0 H_2^0
- H_1^0 H_1^0 H_2^0\right) + \ \ \cdots \ . \eqno(24)
$$
The full expression will be given elsewhere.

%%%%%%%%%%%%%%%%%%%%%%%%%%%%
\vglue.1in

{\bf (C) \ Interactions with fermions}

Our Higgs particles $\phi_L$ do not have interactions with light fermions
$f$ at tree level, and they can couple only between light fermions $f$ and
heavy fermions $F$.
However, when the $6\times 6$ fermion mass matrix  is diagonalized as (9),
the interactions of $\phi_L$ with
the physical fermion states (mass eigenstates) become
$$
\left( \begin{array}{cc}
0 & \Gamma_L \\
0 & 0
\end{array} \right) \Longrightarrow
\left( \begin{array}{cc}
\Gamma_{11} & \Gamma_{12} \\
\Gamma_{21} & \Gamma_{22}
\end{array} \right) \ ,  \eqno(25)
$$
where $\Gamma_L=y_f \phi_L$, and
$$
\Gamma_{11}\simeq U_L^f \phi_L v^{-1} U_L^{f\dagger} D_f \ . \eqno(26)
$$

For charged leptons, since $U_L^e={\bf 1}$,
the interactions of $\phi_L^0$ are given by
$$
H_{Yukawa}^{lepton}=\frac{1}{2\sqrt{2}}\sum_{i,j}\left[
\overline{e}_i(a_{ij}-b_{ij}\gamma_5)e_j (H^0)_i^j
+ i
\overline{e}_i(b_{ij}-a_{ij}\gamma_5)e_j (\chi^0)_i^j
\right] \ ,\eqno(27)
$$
$$
a_{ij}=\frac{m_i}{v_i}+\frac{m_j}{v_j} \ , \ \ \
b_{ij}=\frac{m_i}{v_i}-\frac{m_j}{v_j} \ . \eqno(28)
$$
Therefore, in the pure leptonic modes, the exchange of $\phi_L$ cannot
cause family-number non-conservation.

For quarks, in spite of $U_L^q\neq {\bf 1}$, the Higgs boson $H_1^0$ still
couples with quarks $q_i$ diagonally:
$$
H_{Yukawa}^{quark}=
\frac{1}{\sqrt{2}} \sum_i \frac{m_i^q}{v_0}(\overline{q}_i q_i) H_1^0
\ + \ \ \cdots \ \ . \eqno(29)
$$
However, the dotted parts which are interaction terms of $\phi_2$, $\phi_3$
and $\phi_i^j$ ($i\neq j$) cause family-number non-conservation.

%%%%%%%%%%%%%%%%%%%%%%%%%%%%%%%%%%%%%%%%%%%%%%%%%%%%%% 6.

\section{Family-number changing and conserving neutral currents}

{\bf (A) \  Family-number changing neutral currents}

In general, the Higgs boson $H_1^0$ do not contribute to
flavor-changing neutral currents (FCNC), and only
the other bosons contribute to $\overline{P}^0$-$P^0$
mixing.
The present experimental values [3]
$\Delta m_K =m(K_L)-m(K_S)=(0.5333\pm 0.0027)\times 10^{10}
\ \hbar {\rm s}^{-1}$,
$|\Delta m_D| =|m(D_1^0)-m(D_2^0)|<20\times 10^{10}
\ \hbar {\rm s}^{-1}$,
$\Delta m_B =m(D_H)-m(D_L)=(0.51\pm 0.06)\times 10^{12}
\ \hbar {\rm s}^{-1}$, and so on, give the lower bound of Higgs bosons
$m(H_2^0),\ m(\chi_2^0) > 10^5$ GeV.
For the special case of $m(H)=m(\chi)$,
we obtain the effective Hamiltonian
$$
H_{FCNC}=\frac{1}{3}\left(
\frac{1}{m^2(H_2^0)}-\frac{1}{m^2(H_3^0)}\right)
\sum_{i\neq j}\frac{m_im_j}{v_0^2}
\sum_k \left(
\frac{1}{z_k^2}+\frac{z_k-z_l-z_m}{z_1z_2z_3}\right)
$$
$$
\times
(U_i^kU_j^{k*})^2
\left[(\overline{f}_if_j)^2-(\overline{f}_i\gamma_5f_j)^2\right]
\ , \eqno(30)
$$
where $(k,l,m)$ are cyclic indices of $(1,2,3)$,
so that the bound can reduce to
$m(H_2^0)=\ m(\chi_2^0) >$ a few TeV.
Note that FCNC can highly be suppressed if $m(H_2)\simeq m(H_3)$.

%%%%%%%%%%%%%%%%%%%%%%%%%%%%%%%%%%%%%%%%%
\vglue.1in

{\bf (B) \ Family-number conserving neutral currents}

The strictest restriction on the lower bound of the Higgs boson masses
comes from
$$
\frac{B(K_L\rightarrow e^\pm \mu^\mp)}{B(K_L \rightarrow\pi^0\ell^\pm\nu)}
\simeq
\left(\frac{v_0}{m_H}\right)^4 \times 1.94 \times 10^{-6} \ .\eqno(31)
$$
The present data [3] $B(K_L\rightarrow e^\pm\mu^\mp)_{exp}
< 3.3\times10^{-11}$ leads to the lower bound
$m_{H3}/v_0 > 12$, i.e., $m_{H3} > 2.1$ TeV.

%%%%%%%%%%%%%%%%%%%%%%%%%%%%%%%%%%%%%%%%%%%%%%%%%%%%%% 7.

\section{Productions and decays of the Higgs bosons}

As stated already, as far as our Higgs boson $H_1^0$ is concerned,
it is hard to distinguish it from $H^0$ in the standard model.
We discuss what is a new physics expected concerned with
the other Higgs bosons.

\vglue.1in

{\bf (A) \ Productions}

Unfortunately, since masses of our Higgs bosons $\phi_2$ and $\phi_3$ are
of the order of a few TeV, it is hard to observe a production
$$
e^+ + e^- \rightarrow Z^* \rightarrow (H^0)_i^j \ \ \ \ \ \
+ \ (\chi^0)_j^i \ ,
$$
$$\hspace{2.8cm}\hookrightarrow f_i+\overline{f}_j \ \ \ \ \ \
\hookrightarrow f_j+\overline{f}_i \ , \eqno(32)
$$
even in  $e^+e^-$ super linear colliders which are
planning in the near future.
Only a chance of the observation of our Higgs bosons $\phi_i^j$ is
in a production
$$ u \rightarrow t + (\phi)_1^3  \ ,\eqno(33) $$
at a super hadron collider with several TeV beam energy,
for example, at LHC,
because the coupling  $a_{tu}$ ($b_{tu}$) is sufficiently large:
$$
a_{tu} \simeq \frac{m_t}{v_3} +\frac{m_u}{v_1}=1.029+ 0.002
, \eqno(34)
$$
[c.f. $a_{bd} \simeq ({m_b}/{v_3})+({m_d}/{v_1})
=0.026+ 0.003$].

\vglue.1in

{\bf (B) \ Decays}

Dominant decay modes of $(H^0)_3^{2}$ and $(H^0)_3^{1}$
are hadronic ones,
i.e., $(H^0)_3^{2}\rightarrow t\overline{c}$,
$b\overline{s}$ \ and $(H^0)_3^{1}\rightarrow t\overline{u}$,
$b\overline{d}$ .
Only in $(H^0)_2^{1}$ decay,
a visible branching ratio of leptonic decay is expected:
$$
\Gamma(H_2^{1}\rightarrow c\overline{u}):
\Gamma(H_2^{1}\rightarrow s\overline{d})
:\Gamma(H_2^{1}\rightarrow \mu^-e^+)
$$
$$
\simeq 3\left[\left(\frac{m_c}{v_2}\right)^2+
\left(\frac{m_u}{v_1}\right)^2\right] :
3\left[\left(\frac{m_s}{v_2}\right)^2
+\left(\frac{m_d}{v_1}\right)^2\right] :
\left[\left(\frac{m_{\mu}}{v_2}\right)^2
+\left(\frac{m_e}{v_1}\right)^2\right]
$$
$$
= 73.5\% : 24.9\% : 1.6\% .\eqno(35)
$$

%%%%%%%%%%%%%%%%%%%%%%%%%%%%%%%%%%%%%%%%%%%%%%%%%%%%%% 8

\section{Summary}

We have proposed a U(3)-family nonet Higgs boson scenario,
which leads to a seesaw-type quark and lepton mass matrix
$M_f \simeq m_L M_F^{-1} m_R$.

It has been investigated what a special form of the the potential
$V(\phi)$ can provide the relation
$$
 m_e+m_\mu+m_\tau
=\frac{2}{3}(\sqrt{m_e}+\sqrt{m_\mu}+\sqrt{m_\tau})^2 \ ,
$$
and the lower bounds on the masses of $\phi_L$ have been estimated
from the data of $P^0$-$\overline{P}^0$ mixing and rare meson decays.

Unfortunately, the Higgs bosons, except for $H_1^0$, in the present
scenario are very heavy, i.e., $m_{H}\simeq m_\chi \sim$ a few TeV.
We expect that our Higgs boson $(\phi^0)_1^3$ will be observed through
the reaction $u \rightarrow t+(\phi^0)_1^3$ at LHC.

The present scenario is not always satisfactory from the theoretical
point of view:

\noindent
(1) A curious ansatz, the ``nonet" ansatz, has been assumed.

\noindent
(2) The potential includes an explicitly symmetry breaking term $V_{SB}$.

These problems are future tasks of our scenario.

%%%%%%%%%%%%%%%%%%%%%%%%%%%%%%%%%%%%%%%%%%%%%%%%%%%%%%%%%%%%%%%
%%%%%%%%%%%%%%%%%%%%%%%%%%%%%%%%%%%%%%%%%%%%%%%%%%%%%%%%%%%%%%%%%%%%%
\vglue.3in

\centerline{\bf Acknowledgments}

Portions of this work (quark mass matrix phenomenology)
were begun in collaboration with H.~Fusaoka [6].
I would like to thank him for helpful conversations.
The problem of the flavor-changing neutral currents in the present model
was pointed out by K.~Hikasa.
I would sincerely like to thank Professor K.~Hikasa for valuable comments.
An improved version of this work is in preparation in collaboration with
Prof.~M.~Tanimoto.
I am indebted to Prof.~M.~Tanimoto for helpful comments.
I would also like to thank the organizers of this workshop,
especially, Professor R.~Najima for a successful and enjoyable workshop.
This work was supported by the Grant-in-Aid for Scientific Research,
Ministry of Education, Science and Culture, Japan (No.06640407).

%%%%%%%%%%%%%%%%
\vglue.3in
\newcounter{0000}
\centerline{\bf References and Footnote}
\begin{list}
{[~\arabic{0000}~]}{\usecounter{0000}
\labelwidth=0.8cm\labelsep=.1cm\setlength{\leftmargin=0.7cm}
{\rightmargin=.2cm}}
\item M.~Kobayashi and T.~Maskawa, Prog.~Theor.~Phys. {\bf 49}, 652 (1973).
\item Y.~Koide, Lett.~Nuovo Cimento {\bf 34}, 201 (1982); Phys.~Lett.
{\bf B120}, 161 (1983); Phys.~Rev. {\bf D28}, 252 (1983);
Mod.~Phys.~Lett. {\bf 8}, 2071 (1993).
\item Particle data group, Phys.~Rev. {\bf D50}, 1173 (1994).
\item J.~R.~Patterson, a talk presented at the International Conference on
{\it High Energy Physics}, Glasgow, July 20 -- 27, 1994.
\item Y.~Koide, Mod.~Phys.~Lett. {\bf A5}, 2319 (1990).
\item Y.~Koide and H.~Fusaoka, US-94-02, 1994 (hep-ph/9403354),
(unpublished);
H.~Fusaoka and Y.~Koide, AMUP-94-09 \& US-94-08, 1994 (hep-ph/9501299),
to be published in Mod.~Phys.~Lett. (1995).

Also see, Y.~Koide, Phys.~Rev. {\bf D49}, 2638 (1994).
\item The ``family symmetry" is also called a ``horizontal symmetry":
K.~Akama and H.~Terazawa, Univ.~of Tokyo, report No.~257 (1976)
(unpublished); T.~Maehara and T.~Yanagida, Prog.~Theor.~Phys. {\bf 60},
822 (1978); F.~Wilczek and A.~Zee, Phys.~Rev.~Lett. {\bf 42}, 421 (1979);
A.~Davidson, M.~Koca and K.~C.~Wali, Phys.~Rev. {\bf D20}, 1195 (1979);
J.~Chakrabarti, Phys.~Rev. {\bf D20}, 2411 (1979).
\item Y.~Koide and H.~Fusaoka, US-94-02, (1994), (hep-ph/9403354),
(unpublished).
\end{list}
%%%%%%%%%%%%%%%

\end{document}